\documentclass[12pt]{article}
\usepackage{stmaryrd}
\usepackage{bbm}
\usepackage{mathrsfs}
\usepackage{amssymb}
\usepackage{amsmath}
\usepackage{amsthm}
\usepackage{graphicx}
\usepackage{color}
\usepackage{ulem}
\usepackage[square,numbers,sort&compress]{natbib}

\newtheorem{theorem}{Theorem}

\newtheorem{lemma}{Lemma}

\allowdisplaybreaks[4]
\begin{document}

\title{On realizations of the Virasoro algebra\thanks{Supported by
the NSF of China (Grant Nos. 11101332, 11201371), the Foundation of
Shaanxi Educational Committee, China (Grant No. 11JK0482) and the
NSF of Shaanxi Province, China (Grant No. 2012JQ1013). } }

\author{{ Renat Zhdanov$^{1}$ and Qing Huang$^{2,3}$}\\
{\small 1. BIO-key International, 55121 Eagan, MN, USA}\\
{\small 2. Department of Mathematics, Northwest University, Xi'an 710069, China}\\
{\small 3. Center for Nonlinear Studies, Northwest University, Xi'an 710069, China}}
\date{}

\maketitle
\begin{abstract}
We obtain complete classification of inequivalent
realizations of the Virasoro algebra by Lie vector
fields over the three-dimensional field of real
numbers. As an application we construct new
classes of nonlinear second-order partial
differential equations possessing
infinite-dimensional Lie symmetries.
\end{abstract}

\section{Introduction}

Since its introduction in 19th century Lie group analysis has
become a very popular and powerful tool for solving
nonlinear partial differential equations (PDEs). Given a PDE
that possesses nontrivial symmetry, we can utilize symmetry
reduction procedure to construct its exact solutions \cite{fus93,zhd97}.
Not surprisingly, the wider symmetry of an equation under study
is, the better off we are when applying Lie approach to solve it. This is
especially the case when symmetry group is infinite-parameter. If a
nonlinear differential equation admits infinite Lie symmetry then it
is often possible either to linearize it or construct its general solution
\cite{fus93}.

The classical example is the hyperbolic type Liouville equation
\begin{equation}
\label{lio}
u_{tx}=\exp(u).
\end{equation}
It admits the infinite-parameter Lie group
\begin{equation}
\label{inf}
t'=t+f(t),\quad x'=x+g(x),\quad u'=u-\dot f(t)-\dot g(x),
\end{equation}
where $f$ and $g$ are arbitrary smooth functions. General solution
of (\ref{lio}) can be obtained by the action of transformation group
(\ref{inf}) on its particular traveling wave solution
of the form $u(t,x)=\varphi(x+t)$ (see, e.g., \cite{fus93}). An alternative
way to solve the Liouville equation is to linearize it \cite{fus93}.

Note that the Lie algebra of Lie group (\ref{inf}) is the direct
sum of two infinite-dimensional Witt algebras, which are subalgebras
of the Virasoro algebra.

Unlike the finite-dimensional algebras, infinite-dimensional
ones have not been systematically studied within the
context of classical Lie group analysis of nonlinear PDEs.
The situation is however drastically different for the
case of generalized (higher) Lie symmetries which played critical
role in success of the theory of integrable systems in $(1+1)$ and
$(1+2)$ dimensions (see, e.g. \cite{ibr85}).

The breakthrough in the analysis of integrable systems was nicely
complemented by development of the theory of infinite-dimensional
Lie algebras such as loop algebras \cite{pre86}, Kac-Moody algebras
\cite{kac94} and Virasoro algebras \cite{ioh11}.

Virasoro algebra plays an increasingly important role in
mathematical physics in general \cite{bel84, gra13} and in the theory
of integrable systems in particular. Study of nonlinear
evolution equations in (1+2)-dimensions arising in different
areas of modern physics shows that in many cases the Virasoro algebra
is their symmetry algebra. Let us mention among others the
Kadomtsev-Petvishilvi (KP) \cite{dav85, dav86,gun02}, modified KP,
cylindrical KP \cite{levi88}, the Davey-Stewartson \cite{cha88, gun06},
Nizhnik-Novikov-Veselov, stimulated Raman scattering, (1+2)-dimensional
Sine-Gordon \cite{sen98} and the KP hierarchy \cite{orl97} equations.

Note that there exist integrable equations which admit infinite-dimensional
symmetry algebras that are not of Virasoro type. For instance, the
breaking soliton and Zakharov-Strachan equations do not possess
Virasoro type symmetry while being integrable \cite{sen98}.

It is a common belief that nonlinear PDEs admitting symmetry algebras
of Virasoro type are prime candidates for the roles of integrable systems.
Consequently, systematic classification of inequivalent realizations of
the Virasoro algebra is a crucial step of symmetry approach to 
constructing integrable systems (see, e.g., \cite{lou95,lou04}).

Classification of Lie algebras of vector fields of differential
operators within action of local diffeomorphism group has been
pioneered by Sophus Lie himself. It remains a very powerful method
for group analysis of nonlinear differential equations.
Some of more recent applications of this approach include geometric
control theory \cite{jur97}, theory of systems of nonlinear
ordinary differential equations possessing superposition principle
\cite{shn84}, algebraic approach to molecular dynamics \cite{alh84, sal10} to
mention only a few. Still the biggest bulk of results has been
obtained in the area of classification of nonlinear PDEs possessing
point and higher Lie symmetries (see \cite{zhd01} and references
therein). Analysis of realizations of Lie algebras by first-order
differential operators is in the core of almost every approach to
group classification of partial differential equations (see, e.g.,
\cite{ack75, cam66,gon92a,kam89,kos83,gon92,zhd97})

In this paper we concentrate on study of realizations of Virasoro
algebras by first-order differential operators (Lie vector fields)
in the space $\mathbb{R}^n$ with $n\le 3$. One of our
motivations was to utilize these realizations for constructing
$(1+1)$-dimensional PDEs that possess infinite-dimensional
Lie symmetry and in this sense are integrable.

The paper is organized as follows. In Section 2 we give a brief
account of necessary facts and definitions. In addition, the
algorithmic procedure for classifying realizations
of infinite-dimensional Virasoro and Witt algebras is
described in detail. We construct all inequivalent
realizations of the Witt algebra (a.k.a. centerless
Virasoro algebra) in Section 3. The next section
is devoted to analysis of realizations of the full Virasoro
algebra. We prove that there are no central extensions of the Witt
algebra in the space $\mathbb{R}^3$ that possess nonzero central
element. In Section 5 we construct broad classes of nonlinear
PDEs admitting infinite dimensional symmetry algebras which are
realizations of the Witt algebra. The last section contains
discussion of the obtained results and the outline of further
work.

\section{Notations and definitions}

The Virasoro algebra, $\mathfrak{V}$, is the infinite-dimensional Lie
algebra with basis elements $C, L_n$, $n=0,\pm1,\pm2,\ldots$ which
satisfy the following commutation relations:
\begin{equation}
\label{virasoro} [L_m,\ L_n] = (m-n)
L_{m+n}+\frac{1}{12}m(m^2-1)\delta_{m,-n}C,\quad [L_m,C]=0,\quad
m,n\in\mathbb{Z}.
\end{equation}
Hereafter we denote the commutator of two Lie vector fields $P$ and
$Q$ as $[Q, P]$, i.e., $[Q, P]=QP-PQ$. The symbol $\delta_{a,b}$
stands for the Kronecker delta
\begin{equation*}
\delta_{a,b}=\left\{\begin{array}{ll}1, &a=b,\\[2mm]
    0, & {\rm otherwise}.\end{array}\right.
\end{equation*}
The operator $C$ commuting with all other elements is called the
central element of algebra \eqref{virasoro}. In the case when $C$
equals to zero the algebra \eqref{virasoro} is called the centerless
Virasoro algebra or Witt algebra $\mathfrak{W}$. Consequently, the
full Virasoro algebra is the universal central extension of the Witt
algebra.

The realization space of the Virasoro algebra is the
infinite-dimensional Lie algebra $\mathfrak{L}_{\infty}$ of
first-order differential operators of the form
\begin{equation}
\label{basis}
Q=\tau(t,x,u)\partial_t+\xi(t,x,u)\partial_x+\eta(t,x,u)\partial_u
\end{equation}
over the space $\mathbb{R}^3\ni (t,x,u)$. One can readily verify
that the set of operators (\ref{basis}) is invariant under the
transformation of variables $t,x,u$
\begin{equation}
\label{tr}
t\to \tilde{t}=T(t,x,u),\quad x\to \tilde{x}=X(t,x,u),\quad u\to
\tilde{u}=U(t,x,u),
\end{equation}
provided $D(T,X,U)/D(t,x,u)\neq0$. Indeed, applying \eqref{tr} to an
arbitrary element of $\mathfrak{L}_{\infty}$ of the form
\eqref{basis} we get
\begin{equation*}
\tilde{Q}=(\tau T_t+\xi T_x+\eta T_u)\partial_{\tilde{t}}+(\tau
X_t+\xi X_x + \eta X_u)\partial_{\tilde{x}}+(\tau U_t+\xi U_x+\eta
U_u)\partial_{\tilde{u}}.
\end{equation*}
Evidently, $\tilde{Q}\in\mathfrak{L}_{\infty}$.

It is a common knowledge that correspondence $Q\sim \tilde{Q}$ is the
equivalence relation and as such it splits the set of operators
(\ref{basis}) into equivalence classes. Any two elements within the
equivalence class are related through a transformation (\ref{tr}),
while two elements belonging to different equivalence classless
cannot be transformed one into another by a transformation of the
form (\ref{tr}). Hence to describe all possible realizations of the
Virasoro algebra within the class of Lie vector fields (\ref{basis})
one needs to construct a representative of each equivalence class.
The remaining realizations are obtained by applying
transformations (\ref{tr}) to the representatives 
in question.

The procedure for constructing realizations of inequivalent
algebra \eqref{virasoro} consists of the following three steps:
\begin{itemize}
\item Describe all inequivalent forms of $L_0$, $L_1$ and $L_{-1}$
such that the commutation relations of the Virasoro subalgebra,
\begin{equation} \label{sub}
[L_0,L_1]=-L_1,\quad
[L_0,L_{-1}]=L_{-1},\quad [L_1,L_{-1}]=2L_0,
\end{equation}
hold together with the relations $[L_i,C]=0,\ (i=0,1,-1)$.
Note that the algebra \eqref{sub} is isomorphic to $sl(2,\mathbb{R})$.
\item Construct all inequivalent realizations of the operators $L_2$
and $L_{-2}$ which satisfy the following commutation relations:
\begin{equation}\label{5d}
\begin{split}
&[L_0,L_2]=-2L_2,\quad [L_{-1},L_2]=-3L_1,\quad [L_1,L_{-2}]=3L_{-1},\\[2mm]
&[L_0, L_{-2}]=2L_{-2},\quad[L_2,L_{-2}]=4L_0+\frac12C,\quad [L_i,C]=0,\ (i=2,-2).
\end{split}
\end{equation}
\item Describe the remaining basis operators of the Virasoro algebra
through the recursion relations
\begin{equation*}
L_{n+1}=(1-n)^{-1}[L_1,\ L_n],\quad L_{-n-1}=(n-1)^{-1}[L_{-1},\
L_{-n}]
\end{equation*}
and
\begin{equation*}
   [L_{n+1},L_{-n-1}]=2(n+1)L_0+\frac{1}{12}n(n+1)(n+2)C,\quad [L_i,C]=0
\end{equation*}
with $i=n+1,-n-1$ and $n=2,3,4,\ldots$.
\end{itemize}

In the Sections 3 and 4 we implement the above algorithm to construct
all inequivalent realizations of the Witt and Virasoro
algebras by operators (\ref{basis}).

\section{Realizations of the Witt algebra}
Turn now to describing realizations of the Witt algebra. We remind that
the Witt algebra is obtained from the Virasoro algebra by putting $C=0$.
We begin by letting the vector field $L_0$ be of the general form
\eqref{basis},
i.e.,
\begin{equation*}
L_0=\tau(t,x,u)\partial_t+\xi(t,x,u)\partial_x+\eta(t,x,u)\partial_u.
\end{equation*}
Transformation \eqref{tr} maps $L_0$ into
\begin{equation*}
\tilde{L}_0=(\tau T_t+\xi T_x+\eta T_u)\partial_{\tilde{t}}+(\tau
X_t+\xi X_x+\eta X_u)\partial_{\tilde{x}}+(\tau U_t+\xi U_x+\eta
U_u)\partial_{\tilde{u}}.
\end{equation*}
We have $\tau^2+\xi^2+\eta^2\neq0$, since otherwise $L_0$ is zero.
Consequently, we can choose the solutions of equations
\begin{equation*}
\tau T_t+\xi T_x+\eta T_u=1,\quad\tau X_t+\xi X_x+\eta X_u=0,\quad
\tau U_t+\xi U_x+\eta U_u=0.
\end{equation*}
as $T$, $X$ and $U$ and reduce $L_0$ to the form $L_0=\partial_t$ 
(hereafter we drop the tildes). Consequently, the vector field $L_0$ is
equivalent to the canonical operator $\partial_t$.

With $L_0$ in hand we now proceed to constructing $L_1$ and $L_{-1}$
which obey the commutation relations \eqref{sub}. Let $L_1$ be of
the general form \eqref{basis}. Inserting it into $[L_0,L_1]=-L_1$
yields
\begin{equation*}
L_1=\mathrm{e}^{-t}f(x,u)\partial_t+\mathrm{e}^{-t}g(x,u)\partial_x
+ \mathrm{e}^{-t}h(x,u)\partial_u,
\end{equation*}
where $f,\ g,\ h$ are arbitrary smooth functions. To further
simplify vector field $L_1$ we utilize the equivalence transformation
\eqref{tr} preserving the form of $L_0$. Applying \eqref{tr} to
$L_0$ gives
\begin{equation*}
L_0\to\tilde{L}_0=T_t\partial_{\tilde{t}}+X_t\partial_{\tilde{x}}
+U_t\partial_{\tilde{u}}=\partial_{\tilde{t}}.
\end{equation*}
Consequently, transformation
\begin{equation*}
\tilde{t}=t+T(x,u),\ \tilde{x}=X(x,u),\ \tilde{u}=U(x,u)
\end{equation*}
is the most general transformation that does not alter the form of
$L_0$. It maps the Lie vector field $L_1$ into
\begin{equation*}
\tilde{L_1}=\mathrm{e}^{-t}(f +gT_x+hT_u)\partial_{\tilde{t}}
+\mathrm{e}^{-t}(gX_x+hX_u)\partial_{\tilde{x}}
+\mathrm{e}^{-t}(gU_x+hU_u)\partial_{\tilde{u}}.
\end{equation*}
To further analyze the above class of realizations of $L_1$ we need
to differentiate between two separate cases $g^2+h^2=0$ and
$g^2+h^2\not=0$.

{\bf Case 1.} If $g^2+h^2=0$, then
$\tilde{L_1}=\mathrm{e}^{-t}f(x,u) \partial_{\tilde{t}}$. Choosing
$\tilde{t}=t-\ln{|f(x,u)|}$ we have $L_1=\mathrm{e}^{-t}\partial_t$.
Let $L_{-1}$ be of the general form \eqref{basis}. Inserting $L_0,
L_1, L_{-1}$ into the commutation relations $[L_0,L_{-1}]=L_{-1}$
and $[L_1,L_{-1}]=2L_0$ yields $L_{-1}=\mathrm{e}^t\partial_t$.

{\bf Case 2.} If $g^2+h^2\neq0$ then we choose $\tilde{t}=t+T(x,u)$
where $T(x,u)$ satisfies the equation
\begin{equation*}
\mathrm{e}^{-T}=f+gT_x+hT_u
\end{equation*}
and take solutions of the equations
\begin{equation*}
gX_x+hX_u=\mathrm{e}^{-T},\quad gU_x+hU_u=0
\end{equation*}
as $X$ and $U$ thus mapping $L_1$ into $\mathrm{e}^{-t}(\partial_t+\partial_x)$.
Choosing $L_{-1}$ in the general form \eqref{basis} and taking into
account commutation relations $[L_0,L_{-1}]=L_{-1}$ and
$[L_1,L_{-1}]=2L_0$ we get
\begin{equation*}
L_{-1}=\mathrm{e}^t(1-\mathrm{e}^{-2x}f_1(u))\partial_t
+\mathrm{e}^t(-1-\mathrm{e}^{-2x}f_1(u)
+\mathrm{e}^{-x}g_1(u))\partial_x
+\mathrm{e}^{t-x}h_1(u)\partial_u,
\end{equation*}
where $f_1,g_1,h_1$ are arbitrary smooth functions.

The transformation
\begin{equation}
\label{tr2}
\tilde{t}=t,\quad \tilde{x}=x+X(u),\quad \tilde{u}=U(u)
\end{equation}
evidently does not alter the form of $L_0,L_1$. Applying (\ref{tr})
to $L_{-1}$ yields
\begin{equation*}
\tilde{L}_{-1}=\mathrm{e}^t(1-\mathrm{e}^{-2x}f_1(u))\partial_{\tilde{t}}
+\mathrm{e}^t(-1-\mathrm{e}^{-2x}f_1(u)+\mathrm{e}^{-x}g_1(u)
+\mathrm{e}^{-x}h_1\dot{X})\partial_{\tilde{x}}
+\mathrm{e}^{t-x}h_1(u)\dot{U}\partial_{\tilde{u}}.
\end{equation*}
Here and after the dot over the symbol stands for derivative of the
corresponding function of one variable.

To complete the analysis we need to consider separately the cases
$f_1(u)\neq0$ and $f_1(u)=0$.

Provided $f_1(u)\neq0$ we can choose
$$
X(u)=-\ln{\sqrt{|f_1(u)|}},\quad
\phi(u)=(g_1(u)+h_1(u)\dot{X}(u))/\sqrt{|f_1(u)|}.
$$
Selecting $U$ t0 be a solution of $h_1(u)\dot{U}=\sqrt{|f_1(u)|}$ if
$h_1\neq0$ or an arbitrary non-constant function if $h_1=0$, we have
\begin{equation*}
L_{-1}=\mathrm{e}^{t}(1+\alpha\mathrm{e}^{-2x})\partial_{t}
+\mathrm{e}^{t}(-1+\alpha\mathrm{e}^{-2x}+\mathrm{e}^{-x}\phi(u))\partial_{x}
+\beta\mathrm{e}^{t-x}\partial_{u},
\end{equation*}
where $\alpha=\pm 1$ and $\beta=0,1$.

The case $f_1(u)=0$ leads to the realization
\begin{equation*}
\tilde{L}_{-1}=\mathrm{e}^t\partial_{\tilde{t}}+\mathrm{e}^t(-1+
\mathrm{e}^{-x}g_1(u)+\mathrm{e}^{-x}h_1(u)\dot{X})\partial_{\tilde{x}}
+\mathrm{e}^{t-x}h_1(u)\dot{U}\partial_{\tilde{u}}.
\end{equation*}
Choosing $X=0$ and $U$ to be a solution of $h_1(u)\dot{U}=1$ if
$h_1\neq0$ or an arbitrary non-constant function if $h_1=0$, we
obtain
\begin{equation*}
L_{-1}=\mathrm{e}^t\partial_{t}+\mathrm{e}^t(-1+\mathrm{e}^{-x}g_1(u))\partial_{x}
+\beta\mathrm{e}^{t-x}\partial_{u}
\end{equation*}
with $\beta=0,1$.

We summarize the results obtained above in the following lemma.
\begin{lemma}\label{3d}
There exist only two inequivalent realizations of the algebra
\eqref{sub}
\begin{eqnarray}
&1.&L_0=\partial_t,\quad L_1={\rm e}^{-t}\partial_t,\quad
L_{-1}={\rm e}^{t}\partial_t;\label{rep1}\\[2mm]
&2.&L_0=\partial_t,\quad L_1={\rm
e}^{-t}(\partial_t+\partial_x),\quad
L_{-1}={\rm e}^{t}(1+\alpha{\rm e}^{-2x})\partial_t\label{rep2}\\[2mm]
&& \quad + {\rm e}^t(-1+\phi(u){\rm e}^{-x}+\alpha{\rm
e}^{-2x})\partial_x + \beta{\rm e}^{(t-x)}\partial_u.\nonumber
\end{eqnarray}
Here $\alpha=0,\pm 1$, $\beta = 0,1$ and $\phi(u)$ is an
arbitrary smooth function.
\end{lemma}

To get a complete description of inequivalent realizations of the
Witt algebra, we need to extend algebras \eqref{rep1} and
\eqref{rep2} by the operators $L_{2}$ and $L_{-2}$ and perform the
last two steps of the classification procedure described in Section
2. We first present the final result and then give the detailed
proof.
\begin{theorem}\label{witt}
There exist at most eleven inequivalent realizations of the Witt
algebra $\mathfrak{W}$ over the space $\mathbb{R}^3$. Below we give
the list of representatives of each equivalence class
$\mathfrak{W}_i$, $i=1,2,\ldots,11$.

\noindent {$\mathfrak{W}_1:$}
\begin{equation*}
L_n={\rm e}^{-nt}\partial_t,
\end{equation*}

\noindent {$\mathfrak{W}_2:$}
\begin{equation*}
    L_n=\mathrm{e}^{-nt}\partial_t+\mathrm{e}^{-nt}\left[n+\frac12n(n-1)\alpha\mathrm{e}^{-x}\right]\partial_x,
\end{equation*}

\noindent {$\mathfrak{W}_3:$}
\begin{equation*}
\begin{split}
    &L_n=\mathrm{e}^{-nt+(n-1)x}\left[\mathrm{e}^{2x}-(n+1)\gamma\mathrm{e}^x+\frac12n(n+1)\gamma^2\right](\mathrm{e}^x-\gamma)^{-n-1}\partial_t\\[2mm]
    &\qquad +\mathrm{e}^{-nt+(n-1)x}\left[n\mathrm{e}^x-\frac12n(n+1)\gamma\right](\mathrm{e}^x-\gamma)^{-n}\partial_x,
\end{split}
\end{equation*}

\noindent {$\mathfrak{W}_4:$}
\begin{equation*}
\begin{split}
  & L_0=\partial_t,\\[2mm]
  & L_1=\mathrm{e}^{-t}\partial_t+\mathrm{e}^{-t}\partial_x,\\[2mm]
  &L_{-1}=\mathrm{e}^t(1+\gamma\mathrm{e}^{-2x})\partial_x+\mathrm{e}^t(-1+\gamma\mathrm{e}^{-2x}+\mathrm{e}^{-x}\phi)\partial_x,\\[2mm]
  & L_2=\mathrm{e}^{-2t}f(x,u)\partial_x+\mathrm{e}^{-2t}g(x,u)\partial_x,\\[2mm]
  &L_{-2}=\mathrm{e}^{2t}\left[1+3\gamma\mathrm{e}^{-2x}-\frac12{\mathrm{e}^{-3x}}\left(6\gamma\phi+\phi^3\pm(4\gamma+\phi^2)^{3/2}\right)\right]\partial_t\\[2mm]
  &\qquad+\mathrm{e}^{2t}\left[-2+3\mathrm{e}^{-x}\phi+6\gamma\mathrm{e}^{-2x}-\frac12{\mathrm{e}^{-3x}}
  \left(6\gamma\phi+\phi^3\pm(4\gamma+\phi^2)^{3/2}\right)\right]\partial_x,\\[2mm]
& L_{n+1}=(1-n)^{-1}[L_1,\ L_n],\quad L_{-n-1}=(n-1)^{-1}[L_{-1},\
L_{-n}],\quad n\ge 2,
\end{split}
\end{equation*}
where
\begin{equation*}
\begin{split}
&f(x,u)=\mathrm{e}^x\left[4\mathrm{e}^{4x}-10\mathrm{e}^{3x}\phi-36\gamma\mathrm{e}^{2x}+2\mathrm{e}^{x}\left(31\gamma\phi+6\phi^3\pm6(4\gamma+\phi^2)^{3/2}\right)\right.\\[2mm]
&\qquad \left.-64\gamma^2-54\gamma\phi^2-9\phi^4\mp9\phi(4\gamma+\phi^2)^{3/2}\right]r^{-1} \\[2mm]
&g(x,u)=\mathrm{e}^x\left[8\mathrm{e}^{4x}-16\mathrm{e}^{3x}\phi-2\mathrm{e}^{2x}(44\gamma+5\phi^2)+2\mathrm{e}^{x}\left(44\gamma\phi+9\phi^3\pm9(4\gamma+\phi^2)^{3/2}\right)\right.\\[2mm]
&\qquad \left.-64\gamma^2-54\gamma\phi^2-9\phi^4\mp9\phi(4\gamma+\phi^2)^{3/2}\right]r^{-1},\\[2mm]
&r=4\mathrm{e}^{5x}-10\mathrm{e}^{4x}\phi-40\gamma\mathrm{e}^{3x}+10\mathrm{e}^{2x}\left(6\gamma\phi+\phi^3\pm(4\gamma+\phi^2)^{3/2}\right)-
10\mathrm{e}^x\left(6\gamma^2+6\gamma\phi^2\right.\\[2mm]
&\qquad\left.+\phi^4\pm\phi(4\gamma+\phi^2)^{3/2}\right)+30\gamma^2\phi
+20\gamma\phi^3+3\phi^5\pm(2\gamma+3\phi^2)(4\gamma+\phi^2)^{3/2},
\end{split}
\end{equation*}

\noindent {$\mathfrak{W}_5:$}
\begin{equation*}
   L_n=\mathrm{e}^{-nt+(n-1)x}(\mathrm{e}^{x}\pm n)(\mathrm{e}^x\pm1)^{-n}\partial_t+n\mathrm{e}^{-nt+(n-1)x}(\mathrm{e}^x\pm1)^{1-n}\partial_x,
\end{equation*}

\noindent {$\mathfrak{W}_6:$}
\begin{equation*}
  L_n=\mathrm{e}^{-nt}\partial_t+\gamma\mathrm{e}^{-nt}[\mathrm{e}^{nx}-(\mathrm{e}^x-\gamma)^n](\mathrm{e}^x-\gamma)^{1-n}\partial_x,\\[2mm]
\end{equation*}

\noindent {$\mathfrak{W}_7:$}
\begin{equation*}
\begin{split}
  & L_0=\partial_t,\\[2mm]
  & L_1=\mathrm{e}^{-t}\partial_t+\mathrm{e}^{-t}\partial_x,\\[2mm]
  &L_{-1}=\mathrm{e}^t(1+\gamma\mathrm{e}^{-2x})\partial_t+\mathrm{e}^t(-1+\gamma\mathrm{e}^{-2x}+\mathrm{e}^{-x}\phi)\partial_x,\\[2mm]
  & L_2=\mathrm{e}^{-2t+x}(\mathrm{e}^x-\phi)(\mathrm{e}^{2x}-\mathrm{e}^x\phi-\gamma)^{-1}\partial_t+
    \mathrm{e}^{-2t+x}(2\mathrm{e}^x-\phi)(\mathrm{e}^{2x}-\mathrm{e}^x\phi-\gamma)^{-1}\partial_x,\\[2mm]
  & L_{-2}=\mathrm{e}^{2t-3x}(\mathrm{e}^{3x}+3\gamma\mathrm{e}^x-\gamma\phi)\partial_t+
    \mathrm{e}^{2t-3x}(2\mathrm{e}^x-\phi)(-\mathrm{e}^{2x}+\mathrm{e}^x\phi+\gamma)\partial_x,\\[2mm]
& L_{n+1}=(1-n)^{-1}[L_1,\ L_n],\quad L_{-n-1}=(n-1)^{-1}[L_{-1},\
L_{-n}],\quad n\ge 2,
\end{split}
\end{equation*}

\noindent {$\mathfrak{W}_8:$}
\begin{equation*}
   L_n=\mathrm{e}^{-nt}\partial_t+\mathrm{e}^{-nt}\left[n-sgn(n)\frac{\gamma}{2}\sum_{j=1}^{|n|-1}j(j+1)\mathrm{e}^{-2x}\right]\partial_x,
\end{equation*}

\noindent {$\mathfrak{W}_9:$}
\begin{equation*}
\begin{split}
    & L_n=\frac{\mathrm{e}^{-nt+(n-1)x}}{(\mathrm{e}^x-1)^{n+2}}\left[(-1+\sum_{j=1}^{|n|-1}(2j+1))n+(2n+1)\mathrm{e}^x-(n+2)\mathrm{e}^{2x}
     +\mathrm{e}^{3x}\right.\\[2mm]
     &\qquad \left.+sgn(n)\frac{\phi}2\sum_{j=1}^{|n|-1}j(j+1)\right]\partial_t+\frac{\mathrm{e}^{-nt+(n-1)x}}{(\mathrm{e}^x-1)^{n+1}}
     \left[-(-1+\sum_{j=1}^{|n|-1}(2j+1))n\right.\\[2mm]
     &\qquad \left.-2n\mathrm{e}^x+n\mathrm{e}^{2x}
     -sgn(n)\frac{\phi}2\sum_{j=1}^{|n|-1}j(j+1)\right]\partial_x,
\end{split}
\end{equation*}

\noindent {$\mathfrak{W}_{10}:$}
\begin{equation*}
   L_n=\mathrm{e}^{-nt}\partial_t+n\mathrm{e}^{-nt}\partial_x+\frac{\mathrm{sgn}(n)}2\sum_{j=1}^{|n|}j(j-1)\mathrm{e}^{-nt-2x}\partial_u,
\end{equation*}

\noindent {$\mathfrak{W}_{11}:$}
\begin{equation*}
    L_n=\mathrm{e}^{-nt}\partial_t+\mathrm{e}^{-nt}\left[n+\frac{\alpha n(n-1)}2\mathrm{e}^{-x}\right]\partial_x+\frac{n(n-1)}2\mathrm{e}^{-nt-x}\partial_u,
\end{equation*}
where $n\in\mathbb{Z}$, $\alpha=0,\pm1$, $\gamma=\pm1$,
$\mathrm{sgn}(\cdot)$ is the standard sign function and
\begin{equation*}
\phi(u)= \left\{
\begin{array}{l}
c,\ c\in\mathbb{R},\\
u.
\end{array}
\right.
\end{equation*}
\end{theorem}

\proof To prove the theorem it suffices to consider the case when
the operators $L_0,L_1,L_{-1}$ are of the form \eqref{rep1} or
\eqref{rep2}.

{\bf Case 1.} If $L_0,L_1,L_{-1}$ are given by \eqref{rep1}, then it
is straightforward to verify that due to \eqref{5d} $L_2$ and
$L_{-2}$ are of the forms
\begin{equation*}
L_2=\mathrm{e}^{-2t}\partial_t,\qquad
L_{-2}=\mathrm{e}^{2t}\partial_t.
\end{equation*}
The remaining basis elements of the Witt algebra are easily obtained
through recursion, which yields $L_n=\mathrm{e}^{-nt}\partial_t,\
n\in\mathbb{Z}$. We arrive at the realization $\mathfrak{W}_1$ from
Theorem \ref{witt}.

{\bf Case 2.} Turn now to the realization \eqref{rep2}. Inserting
$L_0,L_1,L_{-1}$ into the commutation relations
$[L_0,L_{-2}]=2L_{-2}$ and $[L_1,L_{-2}]=3L_{-1}$ and solving the
latter for the coefficients of the operator $L_{-2}$ yield
\begin{equation*}
\begin{split}
   & L_{-2}=\mathrm{e}^{2t}(1+3\alpha\mathrm{e}^{-2x}+\psi_1(u)\mathrm{e}^{-3x})\partial_t
    +\mathrm{e}^{2t}(-2+3\phi(u)\mathrm{e}^{-x}+\psi_2(u)\mathrm{e}^{-2x}\\[2mm]
   &\qquad +\psi_1(u)\mathrm{e}^{-3x})\partial_x+\mathrm{e}^{2t}(3\beta\mathrm{e}^{-x}+
   \psi_3(u)\mathrm{e}^{-2x})\partial_u,
\end{split}
\end{equation*}
where $\psi_1,\psi_2,\psi_3$ are arbitrary smooth functions of $u$.

Utilizing the commutation relations $[L_0,L_2]=-2L_2$ and
$[L_{-1},L_2]=-3L_1$ in a similar fashion we derive the form of the
basis element $L_2$
\begin{equation*}
   L_2=\mathrm{e}^{-2t}f(x,u)\partial_t+\mathrm{e}^{-2t}g(x,u)\partial_x+\mathrm{e}^{-2t}h(x,u)\partial_u,
\end{equation*}
where $f,g,h$ are solutions of the system of three PDEs
\begin{subequations}\label{eq1}
\begin{align}
& -3(\alpha\mathrm{e}^{-2x}+1)f+2\alpha\mathrm{e}^{-2x}g+(\phi\mathrm{e}^{-x}+\alpha\mathrm{e}^{-2x}-1)f_x+\beta\mathrm{e}^{-x}f_u+3=0,\label{1.1}\\[2mm]
&    (1-\phi\mathrm{e}^{-x}-\alpha\mathrm{e}^{-2x})f+(\phi\mathrm{e}^{-x}-2)g-\phi_u\mathrm{e}^{-x}h+(\phi\mathrm{e}^{-x}+\alpha\mathrm{e}^{-2x})g_x\label{1.2}\\[2mm]
&    \qquad +\beta\mathrm{e}^{-x}g_u+3=0,\nonumber\\[2mm]
 &   \beta\mathrm{e}^{-x}f-\beta\mathrm{e}^{-x}g+2(1+\alpha\mathrm{e}^{-2x})h-(\phi\mathrm{e}^{-x}+\alpha\mathrm{e}^{-2x}-1)h_x-\beta\mathrm{e}^{-x}h_u=0.\label{1.3}
\end{align}
\end{subequations}

Inserting $L_2$ and $L_{-2}$ into the commutation relation
$[L_2,L_{-2}]=4L_0$ yields three more PDEs
\begin{equation}\label{eq2}
\begin{split}
 &(4\mathrm{e}^{-3x}\psi_1+12\alpha\mathrm{e}^{-2x}+4)f+(-6\alpha\mathrm{e}^{-2x}-3\mathrm{e}^{-3x}\psi_1)g+h\mathrm{e}^{-3x}\dot{\psi_1}\\[2mm] &\quad +(-\mathrm{e}^{-3x}\psi_1+2-3\phi\mathrm{e}^{-x}-\mathrm{e}^{-2x}\psi_2)f_x+(-\mathrm{e}^{-2x}\psi_3-3\mathrm{e}^{-x}\beta)f_u-4=0,\\[2mm]
 &(2\mathrm{e}^{-3x}\psi_1+6\phi\mathrm{e}^{-x}+2\mathrm{e}^{-2x}\psi_2-4)f+(6\alpha\mathrm{e}^{-2x}+2-3\phi\mathrm{e}^{-x}-\mathrm{e}^{-3x}\psi_1-2\mathrm{e}^{-2x}\psi_2)g\\[2mm]
 &\quad+(\mathrm{e}^{-3x}\dot{\psi_1}+3\dot{\phi}\mathrm{e}^{-x}+\mathrm{e}^{-2x}\dot{\psi_2})h+(-\mathrm{e}^{-3x}\psi_1+2-3\phi\mathrm{e}^{-x}-\mathrm{e}^{-2x}\psi_2)g_x\\[2mm]
 &\quad+(-\mathrm{e}^{-2x}\psi_3-3\beta\mathrm{e}^{-x})g_u = 0,\\[2mm]
 &(2\mathrm{e}^{-2x}\psi_3+6\beta\mathrm{e}^{-x})f+(-3\beta\mathrm{e}^{-x}-2\mathrm{e}^{-2x}\psi_3)g+(6\alpha\mathrm{e}^{-2x}+2\mathrm{e}^{-3x}\psi_1+\mathrm{e}^{-2x}\dot{\psi_3}+2)h\\[2mm]
 &\quad+(-\mathrm{e}^{-3x}\psi_1+2-3\phi\mathrm{e}^{-x}-\mathrm{e}^{-2x}\psi_2)h_x+(-\mathrm{e}^{-2x}\psi_3-3\beta\mathrm{e}^{-x})h_u = 0.
\end{split}
\end{equation}

To determine the forms of $L_2$ and $L_{-2}$ we have to solve Eqs.
\eqref{eq1} and \eqref{eq2}. It is straightforward to verify that
the relation
$$
Q=\mathrm{e}^{-t-4x}[\beta\mathrm{e}^{3x} +
\psi_3\mathrm{e}^{2x}+(\beta\psi_2-\phi\psi_3-3\alpha\beta)\mathrm{e}^x
+\beta\psi_1-\alpha\psi_3]\neq0
$$
is the necessary and sufficient condition for the system of
equations \eqref{eq1} and \eqref{eq2} to have the unique solution in
terms of $f_x$ $f_u$, $g_x$, $g_u$, $h_x$ and $h_u$. By this reason
we need to differentiate between the cases $Q=0$ and $Q\neq0$.

{\bf Case 2.1.} Let $Q=0$ or, equivalently, $\beta=\psi_3=0$. Eqs.
\eqref{eq1} and \eqref{eq2} do not contain derivatives of the
functions $f,\ g,\ h$ with respect to $u$. That is why the
derivatives $f_x,\ g_x,\ h_x$ can be expressed in two different ways
using \eqref{eq1} and \eqref{eq2}. Equating the right-hand sides of
the two expressions for $h_x$ yields
\begin{equation*}
    h\mathrm{e}^x\frac{\mathrm{e}^{4x}-2\phi\mathrm{e}^{3x}-\psi_2\mathrm{e}^{2x}-2\psi_1\mathrm{e}^x+3\alpha^2+\phi\psi_1-\alpha\psi_2}
    {(\mathrm{e}^{2x}-\phi\mathrm{e}^x-\alpha)(2\mathrm{e}^{3x}-3\phi\mathrm{e}^{2x}-\psi_2\mathrm{e}^x-\psi_1)}=0.
\end{equation*}
Hence $h=0$. Similarly the compatibility conditions for the
derivatives $f_x$ and $h_x$ give two linear equations for the
functions $f$ and $g$. The determinant of the obtained system of
linear equations does not vanish. Hence the system in question has
the unique solution $f(x), g(x)$. Computing the derivatives of the
so obtained $f$ and $g$ with respect to $x$ and comparing the result
with the previously obtained expressions for $f_x$ and $g_x$ yield
\begin{equation}
\label{eq10}
    (\psi_2-6\alpha)(\phi^3+2\psi_1+\phi\psi_2)\mathrm{e}^{11x}+F_{10}[x,u]=0,
\end{equation}
and
\begin{eqnarray}
\label{eq11}
    &&(10\phi^3\psi_1-3\alpha\phi^2(3\psi_2-8\alpha)+3\phi\psi_1(2\alpha+3\psi_2)+2(5\psi_1^2
    \\[2mm]
    &&\qquad -4\alpha(2\alpha^2-3\alpha_2+\psi_2^2)))\mathrm{e}^{10x}+F_9[x,u]=0.\nonumber
\end{eqnarray}
Hereafter $F_n[x,u],\ n\in\mathbb{N}$ denotes a polynomial in
$\exp(x)$ of the power less than or equal to $n$. To find $f$ and
$g$ we need to construct the most general form of $\phi$ and
$\psi_i\ (i=1,2,3)$ satisfying Eqs. \eqref{eq10} and \eqref{eq11}.
If \eqref{eq10} holds then at least one of the following equations
$\psi_2=6\alpha$ and $\psi_1=-(\phi^3+\phi\psi_2)/2$ should be
satisfied.

Case 2.1.1. If $\psi_2=6\alpha$ then Eqs. \eqref{eq10} and
\eqref{eq11} hold if and only if
\begin{equation*}
16\alpha^3+3\alpha^2\phi^2-6\alpha\phi\psi_1-\phi^3\psi_1-\psi_1^2=0,
\end{equation*}
whence $\psi_1=(-6\alpha\phi-\phi^3\pm(4\alpha+\phi^2)^\frac32)/2$.

Case 2.1.1.1. Suppose now that $\psi_1=(-6\alpha\phi-\phi^3 -
(4\alpha+\phi^2)^\frac32))/2$. Provided $\alpha=0$ we have either
$\psi_1=0$ or $\psi_1=-\phi^3$. The case $\alpha=\psi_1=0$ leads to
$L_{-1}=\mathrm{e}^{t}\partial_t+\mathrm{e}^{t}(-1+\mathrm{e}^{-x}\phi)\partial_x$.
Making the equivalence transformation $\tilde{x}=x+X(u)$ we can
reduce $\phi$ to one of the forms $a=0,\pm1$ thus getting
\begin{equation*}
f=1,\qquad g=2+a\mathrm{e}^{-x}.
\end{equation*}
Utilizing the recurrence relations of the Witt algebra we arrive at
the realization $\mathfrak{W}_2$.

Provided $\alpha=0$ and $\psi_1=-\phi^3$ we can reduce the function
$\phi$ to the form $b=0,\pm1$ by the equivalence transformation
$\tilde{x}=x+X(u)$. Note that in the case when $b=0$ we have
$\psi_1=0$ which leads to the realization $\mathfrak{W}_2$. The case
$b\not= 0$ gives rise to the following forms of $f$ and $g$:
\begin{equation*}
   f=\frac{\mathrm{e}^x(\mathrm{e}^{2x}-3b\mathrm{e}^x+3b^2)}{(\mathrm{e}^x-b)^3},\qquad
   g=\frac{ \mathrm{e}^x(2\mathrm{e}^{x}-3b)}{(\mathrm{e}^x-b)^2}.
\end{equation*}
Hence we get the realization $\mathfrak{W}_3$.

Provided $\alpha=\pm1$ we have $\psi_1=(-6\alpha\phi-\phi^3-
(4\alpha+\phi^2)^\frac32)/2$ and the realization
$\mathfrak{W}_4$ is obtained.

Case 2.1.1.2. Let the function $\psi_1$ be of the form
$\psi_1=(-6\alpha\phi-\phi^3 +(4\alpha +\phi^2)^\frac32))/2$. If
$\alpha=0$ then we have $\psi_1=0$ or $-\phi^3$. This case has
already been considered when we analyzed the Case 2.1.1.1. If the
relation $\alpha=\pm1$ holds then we get the realization
$\mathfrak{W}_4$.

Case 2.1.2. If $\psi_1=-(\phi^3+\phi\psi_2)/2$ then Eq. \eqref{eq10}
takes the form
\begin{equation*}
(4\alpha+\phi^2)(\psi_2-(4\alpha-5\phi^2)/4)(\psi_2-(2\alpha-\phi^2))\mathrm{e}^{10x}+F_9[x,u]=0.
\end{equation*}
To solve the equation above we need to analyze the following three
sub-cases.

Case 2.1.2.1. Provided $\psi_2=(4\alpha-5\phi^2)/4$, Eqs.
\eqref{eq10} and \eqref{eq11} are satisfied if and only if
\begin{equation*}
4\alpha+\phi^2=0.
\end{equation*}
Consequently $\alpha\leq0$ and $\phi=2b(-\alpha)^\frac12$ with
$b=\pm1$.

If $\alpha=-1$ then $\phi=2b,\psi_1=2b,\psi_2=-6$ and furthermore
\begin{equation*}
f=\frac{\mathrm{e}^x(\mathrm{e}^x-2b)}{(\mathrm{e}^x-b)^2},\qquad g=\frac{2\mathrm{e}^x}{\mathrm{e}^x-b}.
\end{equation*}
The realization $\mathfrak{W}_5$ is obtained.

If $\alpha=0$ and $\phi=\psi_1=\psi_2=0$ then we arrive at the
realization $\mathfrak{W}_2$ with $\alpha=0$.

Case 2.1.2.2. Let $\psi_2=2\alpha-\phi^2$ and suppose that Eqs.
\eqref{eq10} and \eqref{eq11} hold. Provided $\alpha=0$ we can
transform $\phi$ to become $b=\pm1$ (note that the case $b=0$ has
already been considered). Then
\begin{equation*}
    f=1,\quad g=\frac{2\mathrm{e}^x-b}{\mathrm{e}^x-b}
\end{equation*}
which yields the realization $\mathfrak{W}_6$.

Given $\alpha=\pm1$ we have
\begin{equation*}
    f=\frac{\mathrm{e}^x(\mathrm{e}^x-\phi)}{\mathrm{e}^{2x}-\mathrm{e}^x\phi-b},\qquad
    g=\frac{\mathrm{e}^x(2\mathrm{e}^x-\phi)}{\mathrm{e}^{2x}-\mathrm{e}^x\phi-b},
\end{equation*}
where $b=\pm1$. Since $\phi$ can be reduced to the form $\tilde{u}$
by the equivalence transformation $\tilde{u}=\phi$ with
$\dot{\phi}\neq0$, the realization $\mathfrak{W}_7$ is obtained.

Case 2.1.2.3. If $4\alpha+\phi^2=0$ and Eqs. \eqref{eq10} and
\eqref{eq11} holds, then we get $\alpha\leq0$, whence $\alpha=0,-1$.

Given the relation $\alpha=0$ we can transform $\phi$ to become
$a=0,\pm1$. Thus $f=1$ and $g=2-a\mathrm{e}^{-x}$ which give rise to
the realization $\mathfrak{W}_8$.

In the case when $\alpha=-1$ we have
\begin{equation*}
      f=\frac{\mathrm{e}^x(4+5\mathrm{e}^x-4\mathrm{e}^{2x}+\mathrm{e}^{3x}+\psi_2)}{(\mathrm{e}^x-1)^4},\quad
      g=\frac{\mathrm{e}^x(-4-4\mathrm{e}^x+2\mathrm{e}^{2x}-\psi_2)}{(\mathrm{e}^x-1)^3}.
\end{equation*}
What is more, the function $\psi_2$ can be reduced to the form
$\tilde{u}$ by the equivalence transformation $\tilde{u}=\psi_2$
provided $\psi_2$ is a nonconstant function. As a result we get the
algebra $\mathfrak{W}_9$.

Summarizing we conclude that the case $Q=0$ leads to the
realizations $\mathfrak{W}_i,\ i=2,3,\cdots,9$.

Turn now to the case $Q\neq0$.

{\bf Case 2.2.} If $Q\neq0$ or, equivalently,
$\beta^2+\psi_3^2\neq0$ then Eqs. \eqref{eq1} and \eqref{eq2} can be
solved with respect to $f_x$, $f_u$, $g_x$, $g_u$, $h_x$ and $h_u$.
The compatibility conditions
\begin{equation*}
   f_{xu}=f_{ux},\quad g_{xu}=g_{ux},\quad h_{xu}=h_{ux}
\end{equation*}
can be rewritten in the form of the system of three linear equations
for the functions $f, g, h$
\begin{equation*}
\begin{split}
 &    a_1f+a_2g+a_3h+d_1=0,\\[2mm]
 &    b_1f+b_2g+b_3h+d_2=0,\\[2mm]
 &    c_1f+c_2g+c_3h+d_3=0.
\end{split}
\end{equation*}
Here $a_i,\ b_i,\ c_i,\ d_i,\ (i=1,2,3)$ are functions of $t,\ x,\
\phi,\ \psi_1,\ \psi_2,\ \psi_3$.

It is straightforward to verify that the system above has the unique
solution $f,\ g,\ h$ when $\beta^2+\psi_3^2\neq0$. We do not present
here the explicit formulae for these functions as they are very
cumbersome. Inserting $f,\ g,\ h$ into Eq. \eqref{1.1} yields
\begin{equation*}
    \alpha\beta^6\mathrm{e}^{42x}+F_{41}[x,u]=0.
\end{equation*}
Consequently we have $\alpha=0$ or $\beta=0$.

Case 2.2.1. If $\beta=0$ then Eq. \eqref{1.1} takes the form
\begin{equation*}
    \alpha\psi_3^6\mathrm{e}^{36x}+F_{35}[x,u]=0,
\end{equation*}
which gives $\alpha=0$ and $\psi_3\neq0$ (since $Q=0$ otherwise).
Taking into account these relations we rewrite Eq. \eqref{eq2} as
follows
\begin{equation*}
\begin{split}
&   \psi_1\psi_3^6\mathrm{e}^{36x}+F_{35}[x,u]=0,\\[2mm]
&  (15\phi^2+2\psi_2)\psi_3^6\mathrm{e}^{37x}+F_{36}[x,u]=0,\\[2mm]
&   (57\phi^2-2\psi_2)\psi_3^7\mathrm{e}^{35x}+F_{34}[x,u]=0.
\end{split}
\end{equation*}
Hence we conclude that $\phi=\psi_1=\psi_2=0$. Inserting these
formulae into the initial Eqs. \eqref{eq1} and \eqref{eq2} and
solving the obtained system yield
\begin{equation*}
    f=1,\ g=2,\ h=-\mathrm{e}^{-2x}\psi_3.
\end{equation*}
The function $\psi_3$ can be reduced to the form $-1$ by the
equivalence transformation $\tilde{u}=U(u)$, where
$\dot{U}=-1/\psi_3$. As a result we have
\begin{equation*}
    f=1,\ g=2,\ h=\mathrm{e}^{-2x},
\end{equation*}
which leads to the realization $\mathfrak{W}_{10}$.

Case 2.2.2. Provided $\alpha=0$ Eq. \eqref{1.3} takes the form
\begin{equation*}
    \beta^5(4\beta\phi\psi_3-6\psi_3^2+\beta^2\dot{\psi_3})
    \mathrm{e}^{41x}+30\beta^5\phi\psi_3^2\mathrm{e}^{40x}+F_{39}[x,u]=0.
\end{equation*}
Note that the case $\alpha=\beta=0$ has been already studied while
considering Case 2.2.1. Consequently, without any loss of generality
we can restrict our considerations to the cases $\psi_3=0$,
$\beta=1$ and $\phi=0$, $\beta=1$.

If $\psi_3=0$ then it follows from \eqref{eq2} and \eqref{1.3} that
$\psi_1=\psi_2=0$. Taking into account these relations we rewrite
Eqs. \eqref{eq1} and \eqref{eq2} in the form
\begin{equation*}
    f=1,\ g=2+\mathrm{e}^{-x}\phi,\ h=\mathrm{e}^{-x}.
\end{equation*}
What is more, the function $\phi$ can be reduced to one the forms
$0,\pm1$ by equivalence transformations $\tilde{x}=x+X(u)$ and
$\tilde{u}=U(u)$. Whence we get the realization $\mathfrak{W}_{11}$.

If the relation $\phi=0$ holds Eqs. \eqref{eq1} and \eqref{eq2} are
incompatible.

It is straightforward to verify that $\mathfrak{W}_i,\
(i=1,2,\cdots,11)$ cannot be transformed one into another with the
transformation \eqref{tr} and hence are inequivalent. This completes
the proof of Theorem \ref{witt}.\qed

While proving the above theorem we obtained exhaustive description
of realizations of the Witt algebra over the spaces $\mathbb{R}^1$
and $\mathbb{R}^2$. We present the corresponding results without
proof.
\begin{theorem}\label{witt_r1}
The realization $\mathfrak{W}_1$ exhausts the list of inequivalent
realizations of the Witt algebra $\mathfrak{W}$ over the space
$\mathbb{R}^1$.
\end{theorem}
\begin{theorem}\label{witt_r2}
The realizations $\mathfrak{W}_1$--$\mathfrak{W}_9$ with $\phi=C$ exhaust
 the list of inequivalent realizations of the Witt algebra $\mathfrak{W}$
 over the space $\mathbb{R}^2$.
\end{theorem}

\section{Realizations of the Virasoro algebra}

To construct all inequivalent realizations of the Virasoro algebra,
$\mathfrak{V}$, we need to extend inequivalent realizations of the Witt algebra by
all possible realizations of the nonzero central element of the
Virasoro algebra. In this section we prove that there are no
realizations of the Virasoro algebra with non-zero central element
in the three-dimensional space $\mathbb{R}^3$.

Let us begin by constructing all possible central extensions of the
subalgebra $\mathfrak{L}=\langle L_{-1}, L_0, L_1\rangle$. In view
of Lemma \ref{3d} we can restrict our considerations to realizations
\eqref{rep1} and \eqref{rep2} of the algebra $\mathfrak{L}$.

{\bf Case 1.} Consider first the realization \eqref{rep1}
\begin{equation*}
  L_0=\partial_t,\ L_1=\mathrm{e}^{-t}\partial_t,\
  L_{-1}=\mathrm{e}^t\partial_t.
\end{equation*}
Let the basis element $C$ be of the general form \eqref{basis}.
Inserting \eqref{basis} into the commutation relations $[L_i,C]=0,\
(i=0,1,-1)$ yields
\begin{equation*}
    C=\xi(x,u)\partial_x+\eta(x,u)\partial_u,\quad \xi^2+\eta^2\neq0.
\end{equation*}
Applying the transformation
\begin{equation*}
    \tilde{t}=t,\quad \tilde{x}=X(x,u),\quad \tilde{u}=U(x,u),
\end{equation*}
which does not alter the forms of $L_0$, $L_1$ and $L_{-1}$ to the
realization of $C$ above, we get
\begin{equation*}
    C\to\tilde{C}=(\xi X_x+\eta X_u)\partial_{\tilde{x}}+(\xi U_x+\eta U_u)\partial_{\tilde{u}}.
\end{equation*}
Choosing solutions of the equations
\begin{equation*}
    \xi X_x+\eta X_u=0,\quad\xi U_x+\eta U_u=1,
\end{equation*}
as $X$ and $U$ yields $C=\partial_u$.

Proceed now to constructing $L_2$. It easily follows from the
relations $[L_0,L_2]=-2L_2$, $[L_{-1},L_2]=-3L_1$ and $[L_2,C]=0$
that $L_2=\mathrm{e}^{-2t}\partial_t$. Next, let $L_{-2}$ be of the
general form \eqref{basis}. Then commutation relations \eqref{5d}
involving $L_{-2}$ yield over-determined system of PDEs for the
unknown functions $\tau$, $\xi$ and $\eta$. This system turns out to
be incompatible. Hence realization \eqref{rep1} cannot be extended
to a realization of the Virasoro algebra with nonzero central
element.

{\bf Case 2.} We begin by utilizing commutation relations for the
basis elements $L_0$, $L_1$ and $C$ thus getting
\begin{equation*}
C=f(u)\mathrm{e}^{-x}\partial_t+(g(u)+f(u)\mathrm{e}^{-x})\partial_x+h(u)\partial_u,
\end{equation*}
where $f$, $g$ and $h$ are arbitrary smooth function of $u$.
Applying transformation \eqref{tr2} that preserves the form of the
basis elements $L_0, L_1$ to $C$ gives
 \begin{equation*}
    \tilde{C}=f(u)\mathrm{e}^{-x}\partial_{\tilde{t}}+(g(u)+f(u)\mathrm{e}^{-x}+h(u)\dot{X}(u))\partial_{\tilde{x}}
    +h(u)\dot{U}(u)\partial_{\tilde{u}}.
\end{equation*}
If $f(u)\neq0$ then choosing $X(u)=-\ln{|f(u)|}$ we have
$\tilde{C}=\mathrm{e}^{-\tilde{x}}\partial_{\tilde{t}}+
(\mathrm{e}^{-\tilde{x}}+\beta(g+h\dot{X}))\partial_{\tilde{x}}
+\beta h\dot{U}\partial_{\tilde{u}}$, where $\beta=\pm1$. Provided
$h=0$ and $\dot{g}\neq0$ we can make the transformation $\tilde{u}=g(u)$
and thus get $C_1=\mathrm{e}^{-x}\partial_t +(\mathrm{e}^{-x}+
u)\partial_x$. The case $h=\dot{g}=0$ leads to
$C_2=\mathrm{e}^{-x}\partial_t + (\mathrm{e}^{-x}
+\lambda)\partial_x$, where $\lambda$ is an arbitrary constant.
Next, if $h\neq0$ we choose solutions of the equations
$g+h\dot{X}=0$ and $h\dot{U}=1/\beta$ as $X$ and $U$ getting
$C_3=\mathrm{e}^{-x}\partial_t + \mathrm{e}^{-x}\partial_x +
\partial_u$.

Provided $f(u)=0$ the generator $\tilde{C}=(g +
h\dot{X})\partial_{\tilde{x}} + h\dot{U}\partial_{\tilde{u}}$ is
obtained. If $h\neq0$ then it is possible to choose $X$ and $U$ so
that $C_4=\partial_u$. Given the condition $h=0$ we have
$\tilde{C}=g\partial_{\tilde{x}}$. If $g$ is nonconstant then
selecting $U=g(u)$ yields $C_5=u\partial_x$. Finally, the case of
constant $g$ leads to $C_6=\partial_x$.

Summing up we conclude that there exist six inequivalent nonzero
realizations of the central element $C$ for the case when
$L_0=\partial_t$ and $L_1=\mathrm{e}^{-t}\partial_t +
\mathrm{e}^{-t}\partial_x$. The next step is extending the algebras
$\langle L_0,L_1,C_i \rangle,\ (i=1,2,\cdots,6)$ to the realizations
of the full Virasoro algebra. We present the calculation details for
the case when $C_1=\mathrm{e}^{-x}\partial_t +\mathrm{e}^{-x}
\partial_x +\partial_u$. The remaining five cases are handled in a similar
fashion.

In order to extend $\langle L_0,L_1,C_1 \rangle$ to the full Virasoro
algebra, we construct al  possible realizations of $L_{-1}$. Inserting $L_{-1}$
of the general form \eqref{basis} into the corresponding commutation
relations from \eqref{sub} gives
\begin{equation*}
    L_{-1}=\frac{\mathrm{e}^{t-2x}(u^2\mathrm{e}^{2x}-1)}{u^2}\partial_t-
    \frac{\mathrm{e}^{t-2x}(u\mathrm{e}^{x}+1)^2}{u^2}\partial_x.
\end{equation*}
With $L_{-1}$ in hand we proceed to constructing $L_2$. Taking into account
relations \eqref{5d} we get
\begin{equation*}
    L_2=\frac{u\mathrm{e}^x(u\mathrm{e}^x+2)}{\mathrm{e}^{2t}(u\mathrm{e}^x+1)^2}\partial_t+
    \frac{2u\mathrm{e}^x}{\mathrm{e}^{2t}(u\mathrm{e}^x+1)}\partial_x.
\end{equation*}
Inserting the obtained expressions for $L_{-1}$ and $L_2$ into \eqref{5d} yields
incompatible system of equations for the coefficients of $L_{-2}$. Whence we conclude
that a realization of the algebra $\langle L_0,L_1,C_1 \rangle$ cannot be extended to a
realization of the full Virasoro algebra. The same result holds for the remaining five
realizations of the central elements $C_2,C_3,\ldots,C_6$.

\begin{theorem}
There are no realizations of the Virasoro algebra with nonzero central
element $C$ in three-dimensional space $\mathbb{R}^n,\ n=1,2,3$.
\end{theorem}

\section{Some applications: PDEs admitting infinite-dimensional symmetry
groups}

In this section we construct several classes of second-order evolution equations
in the space $\mathbb{R}^3$ of the variables $t,\ x,\ u$ that admit
realization of the Witt algebra listed in Theorem \ref{witt}. Given a
realization of the Witt algebra, we can apply the Lie infinitesimal approach
to construct the corresponding invariant equation \cite{olv86,ovs82}. Differential
equation
\begin{equation*}
    F(t,x,u,u_t,u_x,u_{tt},u_{tx},u_{xx})=0
\end{equation*}
is invariant with respect to the Witt algebra with basis elements
$L_1,L_2,\ldots,L_n,\ldots$ if and only if the condition
\begin{equation*}
    \text{pr}^{(2)}L_n(F)|_{F=0}=0
\end{equation*}
holds for any $n\in\mathbb{N}$, where $\text{pr}^{(2)}L_n$ is the
second-order prolongation of the vector field $L_n$, that is
\begin{equation*}
\text{pr}^{(2)}L_n=L_n+\eta^t\partial_{u_t}+\eta^x\partial_{u_x}+\eta^{tt}\partial_{u_{tt}}+
\eta^{tx}\partial_{u_{tx}}+\eta^{xx}\partial_{u_{xx}}
\end{equation*}
with
\begin{align*}
        \eta^t      &= D_t(\eta)-u_tD_t(\tau)-u_xD_t(\xi), \\[2mm]
        \eta^x      &= D_x(\eta)-u_tD_x(\tau)-u_xD_x(\xi), \\[2mm]
        \eta^{tt}   &= D_t(\eta^t)-u_{tt}D_t(\tau)-u_{tx}D_t(\xi), \\[2mm]
        \eta^{tx}   &= D_x(\eta^t)-u_{tt}D_x(\tau)-u_{tx}D_x(\xi), \\[2mm]
        \eta^{xx}   &= D_x(\eta^x)-u_{xt}D_x(\tau)-u_{xx}D_x(\xi).
\end{align*}
Here the symbols $D_t$ and $D_x$ stand for the total differentiation
operators with respect to $t$ and $x$, correspondingly
\begin{align*}
  D_t &= \partial_t+u_t\partial_u+u_{tt}\partial_{u_t}+u_{xt}\partial_{u_x}+\dots,
  \\[2mm]
  D_x &= \partial_x+u_x\partial_u+u_{tx}\partial_{u_t}+u_{xx}\partial_{u_x}+\dots.
\end{align*}

As an example we consider the realization $\mathfrak{W}_1 =
\langle\mathrm{e}^{-nt}\partial_t\rangle$. Making use of the formulas
above we obtain
\begin{equation}\label{Q_2}
    \text{pr}^{(2)}L_n=\mathrm{e}^{-nt}\partial_t+n\mathrm{e}^{-nt}u_t\partial_{u_t}
    +(2n\mathrm{e}^{-nt}u_{tt}-n^2\mathrm{e}^{-nt}u_t)\partial_{u_{tt}}
    +n\mathrm{e}^{-nt}u_{tx}\partial_{u_{tx}}.
\end{equation}
Next step is computation of the full set of functionally-independent
second-order differential invariants,
$I_m(t,x,u,u_t,u_x,u_{tt},u_{tx},u_{xx})\ (m=1,2,\cdots,7),$
associated with $L_n$. To get $I_m$ we need to solve the related system of
characteristic equations
\begin{equation*}
    \frac{\mathrm{d}t}{\mathrm{e}^{-nt}}=\frac{\mathrm{d}x}{0}=
    \frac{\mathrm{d}u}{0}=\frac{\mathrm{d}u_t}{n\mathrm{e}^{-nt}u_t}
    =\frac{\mathrm{d}u_x}{0}=\frac{\mathrm{d}u_{tt}}
    {2n\mathrm{e}^{-nt}u_{tt}-n^2\mathrm{e}^{-nt}u_t}
    =\frac{\mathrm{d}u_{tx}}{n\mathrm{e}^{-nt}u_{tx}}
    =\frac{\mathrm{d}u_{xx}}{0}.
\end{equation*}
Integration of the equations above yields
\begin{equation*}
  I_1=x,\ I_2=u,\ I_3=u_x,\ I_4=u_{xx},\ I_5=\frac{u_{tx}}{u_t},\
  I_6=\mathrm{e}^{-nt} u_t,\ I_7=\mathrm{e}^{-2nt}u_{tt}-n\mathrm{e}^{-2nt}u_t.
\end{equation*}
Hence the most general $L_n$-invariant equation is of the form
\begin{equation*}
   F(I_1,I_2,\cdots,I_7)=0.
\end{equation*}
Since this equation has to be invariant under every basis element
of the infinite-dimensional Witt algebra $\mathfrak{W}_1$, it must be
independent of $n$. To obey this restriction function $F$ should be
independent of $I_6$ and $I_7$. Thus the final form of the most general
second-order PDE invariant under $\mathfrak{W}_1$ reads as
\begin{equation*}
   F(I_1,I_2,I_3,I_4,I_5)=0,
\end{equation*}
or,
\begin{equation*}
F\left(x,u,u_x,u_{xx},\frac{u_{tx}}{u_t}\right)=0.
\end{equation*}

Below we list five more classes of second-order differential equations whose
symmetry algebra is infinite-dimensional Witt algebra. 

\noindent $\mathfrak{W}_2$ invariant PDEs
\begin{equation*}
    F\left(u,u_x,u_{xx},\frac{u_tu_{xx}-u_xu_{tx}}{\mathrm{e}^xu_x}\right)=0,\quad {\rm if}\ \alpha=0
\end{equation*}
\begin{equation*}
    F\left(u,\frac{u_{xx}-u_x}{u_x^2},\frac{u_tu_x-u_tu_{xx}+u_xu_{tx}+u_x^2}{\mathrm{e}^xu_x}-2\alpha u_x\right)=0,\quad 
    {\rm if}\ \alpha=\pm1,
\end{equation*}

\noindent $\mathfrak{W}_6$ invariant PDEs
\begin{equation*}
    F\left(u,\frac{\gamma(u_{xx}+u_{tx})-\mathrm{e}^x(u_x+u_{xx})}{u_x(\gamma(u_t+u_x)-\mathrm{e}^xu_x)}\right)=0,
\end{equation*}

\noindent $\mathfrak{W}_8$ invariant PDEs
\begin{equation*}
    F\left(u,\frac{u_{xx}-2u_x}{u_x^2}\right)=0,
\end{equation*}

\noindent $\mathfrak{W}_{10}$ invariant PDEs
\begin{equation*}
    F(u_x+2u,u_{xx}-4u)=0.
\end{equation*}

\section {Concluding Remarks}

The principal result of this paper is exhaustive classification
of inequivalent realizations of the Virasoro algebra by Lie
vector fields over the space $\mathbb{R}^n$ with $n=1,2,3$.
These realizations are listed in Theorems 1-4. According to
Theorem 1 there exist eleven inequivalent realizations of
the Witt algebra in the space $\mathbb{R}^3$. What is more,
we proved that there exist only one realization of the Witt algebra
over the space $\mathbb{R}^1$ and nine realizations of the Witt
algebra over the space $\mathbb{R}^2$.

It has been established that realizations of the Virasoro algebra
with nonzero central element do not exist in the space $\mathbb{R}^n$
with $n\le 3$.

As an application of our algebraic classification we construct
a number of nonlinear PDEs admitting infinite-dimensional symmetry
algebras, which are realizations of the Witt algebra.

An interesting application of the obtained results would be describing
nonlinear PDEs whose symmetry algebras are direct sums of the Witt
algebras. A nontrivial example is the Liouville equation (\ref{lio}).
Since these equations would admit symmetry with two arbitrary
functions they would automatically be classically integrable.

Since Virasoro algebra is a subalgebra of the Kac-Moody algebra, 
the results of this paper can be directly applied to solving
the problem of classification of integrable KP type PDEs in $(1+2)$
dimensions. The starting point is a description of inequivalent
realizations of the Kac-Moody algebras by differential operators
over the space $\mathbb{R}^4$.

These problems are under study now and will be reported in our
future publications.

\end{document}